\documentclass[preprint]{vgtc}

\graphicspath{{figures/}{pictures/}{images/}{./}} 

\usepackage{times}                     
\usepackage{microtype}
\usepackage{amsmath}
\usepackage{enumitem}
\usepackage{float}

\usepackage{mathptmx}                  

\onlineid{xxxx}

\vgtccategory{Research}

\vgtcinsertpkg

\preprinttext{To appear at the IEEE ISMAR 2025 conference.}

\definecolor{blue}{RGB}{0,0,0}

\title{HeadZoom: Hands-Free Zooming and Panning for 2D Image Navigation Using Head Motion}

\author{Kaining Zhang\thanks{e-mail: kaining.zhang@mymail.unisa.edu.au}\\ %
        \scriptsize Empathic Computing Lab, University of South Australia \\ Adelaide, Australia %
\and Catarina Moreira\thanks{e-mail: Catarina.PintoMoreira@uts.edu.au}\\ %
     \scriptsize Data Science Institute, University of Technology Sydney \\ Sydney, Australia %
\and Pedro Belchior\thanks{e-mail: pedro.fls.belchior@inesc-id.pt}\\ %
     \scriptsize INESC-ID, Instituto Superior Técnico da Universidade de Lisboa, \\ Lisboa, Portugal %
\and Gun A. Lee\thanks{email: Gun.Lee@unisa.edu.au} \\
        \scriptsize Empathic Computing Lab, University of South Australia \\ Adelaide, Australia %
\and Mark Billinghurst\thanks{email: Mark.Billinghurst@unisa.edu.au} \\
        \scriptsize Empathic Computing Lab, University of South Australia \\ Adelaide, Australia %
\and Joaquim Jorge\thanks{e-mail: jorgej@tecnico.ulisboa.pt}\\ %
     \scriptsize INESC-ID/Instituto Superior Técnico da Universidade de Lisboa \\ 
     Lisboa, Portugal}

\teaser{
  \centering
 \includegraphics[width=\linewidth]{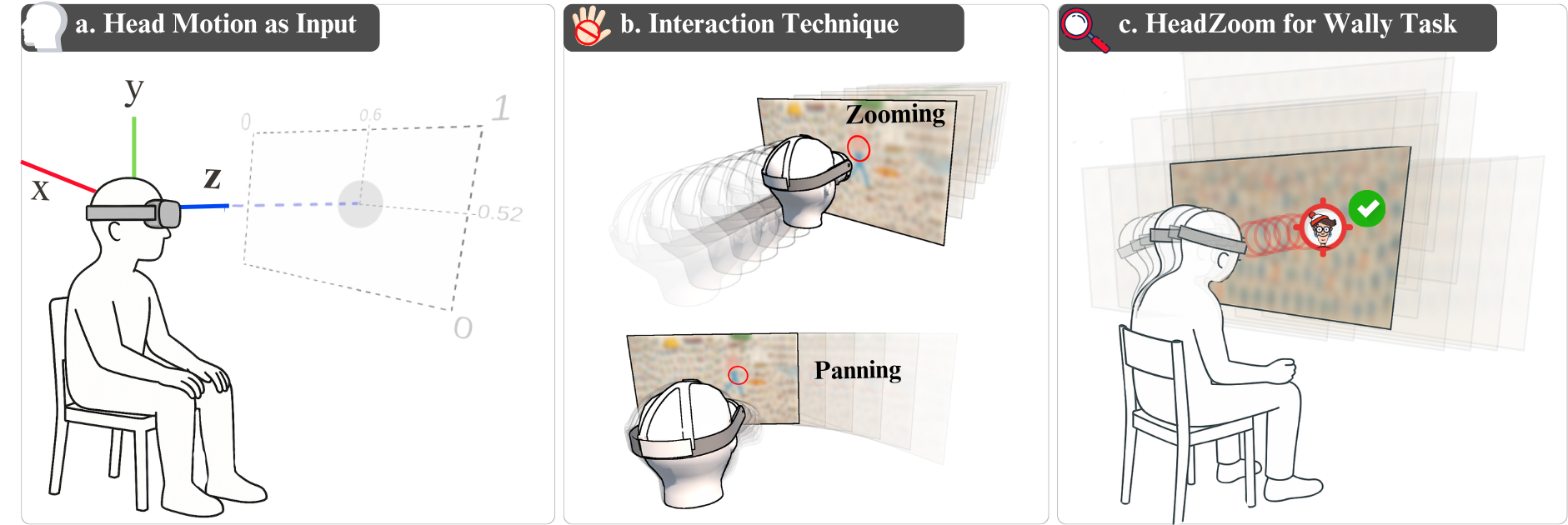}
 \caption{%
HeadZoom enables hands-free, controller-free exploration of 2D imagery using natural head motion. 
Shown here are: (a) the tracked head pose axes used for input; 
(b-top) the Parallel mode, where the image remains front-facing while the user zooms in by leaning forward; 
(b-bottom) the Tilt mode, where zoom is identical, but the image dynamically tilts to stay perpendicular to the user's head; 
(c) a successful identification of the target character (Wally) in a search task.%
}
  \label{fig:teaser}
}
\abstract{
  We introduce \textit{HeadZoom}, a hands-free interaction technique for navigating two-dimensional visual content using head movements. HeadZoom enables fluid zooming and panning using only real-time head tracking. It supports natural control in applications such as map exploration, radiograph inspection, and image browsing, where physical interaction is limited. We evaluated HeadZoom in a within-subjects study comparing three interaction techniques—Static, Tilt Zoom, and Parallel Zoom—across spatial, error, and subjective metrics. Parallel Zoom significantly reduced total head movement compared to Static and Tilt modes. Users reported significantly lower perceived exertion for Parallel Zoom, confirming its suitability for prolonged or precision-based tasks. By minimizing movement demands while maintaining task effectiveness, HeadZoom advances the design of head-based 2D interaction in VR and creates new opportunities for accessible hands-free systems for image exploration.

} 


\begin{document}


\maketitle
\section{Introduction}

 \textcolor{blue}{Zooming and panning interactions are essential for inspecting detailed visual content in high-resolution 2D images \cite{cockburn_surveys09}. Traditional techniques rely on hand-based input, such as a mouse scroll-wheel \cite{igarashi2000speed}, touch-screen gestures \cite{avery2014pinch,mendes14}, 6DOF controllers \cite{digital_slide_navigation}, or shortcut keys \cite{usability_zoom_pan}. However, these methods assume that the user’s hands are free, which may not be true. With the emergence of Mixed Reality (MR), applications have expanded into critical real-world domains of surgery \cite{MR_Surgery}, fieldwork \cite{MR_Fieldwork}, manufacturing \cite{MR_manufacturing}, education \cite{MR_education}, and accessibility \cite{MR_accessibility}. Here, hands are usually occupied with primary tasks, such as gripping instruments, wearing gloves, or maintaining a sterile field \cite{Handsfree_important, Handsfree_neurosurgery}. So using a hand or a controller or touching a surface becomes inconvenient, unsafe, or impossible.}
 
\textcolor{blue}{Current hands-free interaction alternatives have clear limitations. Speech suffers from recognition latency and is unusable in noise-sensitive or silence-required environments \cite{Handsfree_neurosurgery}. Eye-gaze techniques relying on dwell-based selection frequently lead to unintended activations, tracking inaccuracies, and visual fatigue \cite{Handsfree_gaze}. Therefore, there is a need for more natural, precise, and practical hands-free methods. Head movements closely reflect natural user behaviours during zooming and panning operations. People instinctively lean forward to inspect details and rotate their heads to scan a scene, suggesting that head translation and rotation can be intuitive metaphors for magnification and viewport movement. Previous research \cite{yan2018} investigated discrete head-gesture-based zooming, but used fixed zoom increments per gesture, forcing repetitive head motions. This resulted in user fatigue and reduced precision, and notably, the method lacked support for continuous panning operations. There remains a significant research gap regarding continuous, fine-grained, and real-time control of zoom and pan via head movements.}
 
\textcolor{blue}{To address this gap, we introduce \textit{HeadZoom}, a novel interaction framework that maps head translation to zoom and head yaw/pitch rotations to image translation, using adaptive control-display gain to ensure smooth operation. We developed two interaction techniques (Parallel Zoom (PZ), Tilt Zoom (TZ)) and evaluated them against a baseline (Static), which uses natural head movements without specific mapping. Our results indicate that the PZ mode significantly reduces total head movement compared to both Static and TZ. In addition, participants reported substantially lower perceived exertion when using PZ, demonstrating its suitability for extended tasks. By minimizing physical demands while maintaining task performance, \textit{HeadZoom} provides a viable option for hands-free 2D interactions in MR, opening new possibilities for immersive and accessible image exploration. The paper's main contributions include:}


    \noindent \textbf{Head-Based Interaction Framework:} A new method uses head movement to control the zoom and panning of 2D content, ensuring smooth transitions and accurate operation.
    
    \noindent \textbf{User-Centered Design}: An interface that reduces cognitive and physical effort, with efficient interaction in extended use, complemented by three application scenarios and actionable design guidelines to support future practical implementations in MR environments.
    
    \noindent \textbf{Extensive Evaluation}: \textcolor{blue}{A user study comparing the Parallel Zoom and Tilt Zoom techniques against a Static baseline.}

\textcolor{blue}{In the rest of the paper, we first review related work, clarifying the research gap that HeadZoom addresses and our novel contribution. Next, we describe the HeadZoom implementation, before presenting a user study to evaluate the technique. After reporting on the user study results we discuss them in Section~\ref{sec:discussion}. We then describe application scenarios, design guidelines, and limitations. Finally, we close with conclusions and future research directions.}

\section{Related Work}

Navigating visual content often requires \textit{panning} (moving the viewport) and \textit{zooming} (adjusting scale). Zooming and positioning techniques are critical to improving navigation efficiency in different applications. Many software tools, such as image editors or maps, provide interfaces for zooming- and panning-based navigation. This section reviews existing research in these areas and highlights the role of head-based interaction in addressing current challenges.

\subsection{Zooming and Panning Interfaces}
Zooming and panning interfaces have been actively investigated in 2D graphical user interfaces (GUI). 
Perlin and Fox \cite{perlin1993pad, bederson1996padpp} pioneered using zooming and panning as a key method to navigate and manage an infinite 2D information space at various scale levels. They also introduced a concept of semantic zooming that presented a single entity in different representations depending on the zoom level. 

Zooming and panning interfaces have been proposed for the 2D desktop GUI. You et al. \cite{usability_zoom_pan} conducted a usability evaluation of the zoom and pan functions in web-based map interfaces, comparing different zoom and pan button layouts. Gutwin and Fedak~\cite{gutwin2004interacting} used panning and zooming interfaces to fit large 2D GUIs to smaller screens. Igarashi and Hinckley \cite{igarashi2000speed} automatically controlled the zoom level in proportion to the scrolling speed to help users navigate large documents efficiently and smoothly, without getting disoriented. This approach is superior to manual control, which suffers from a fast visual flow when scrolling too quickly while zoomed in.
Cockburn et al.~\cite{cockburn_surveys09}  
provide a comprehensive review of early research on interfaces for navigating large documents in 2D desktop GUI, categorizing them into \textit{Overview+Detail}, \textit{Zooming}, \textit{Focus+Context}, and \textit{Cue-based} techniques. 
Standard input methods for zooming and panning in 2D GUI include mouse, touch gestures, and shortcut keys \cite{usability_zoom_pan}. In addition to these 
popular techniques, researchers also investigated other input methods, such as 6DOF controllers~\cite{digital_slide_navigation} or gaze~\cite{stellmach2012investigating, hansen2008noise}. 

The popularization of mobile devices with small screens further emphasized the need to explore zooming and panning techniques \cite{gutwin2004interacting}. Although the pinch-to-zoom method became widely adopted in touch screen-based mobile devices \cite{avery2014pinch}, the limitation of occupying both hands motivated many researchers to explore a one-handed solution to control zooming and panning~\cite{mendes14}. Researchers have developed various touch gestures \cite{farhad2018evaluating, lai2017contextzoom} to replace or complement the standard pinch-to-zoom gesture. At the same time, others also explored using additional detectors, such as pressure sensors combined with a touch screen to enable natural zooming and scrolling \cite{miyaki2009graspzoom, suzuki2018pressure}, or using a gyro sensor or buttons built into the device \cite{garg2021comparison}.

Zooming and panning interfaces have also been explored in AR and VR. Handheld controllers \cite{chang2017panning} or hand gestures and gaze input \cite{satriadi2019augmented, pfeuffer2017gazepinch, messaci2022zoom} are used to control zoom and pan in VR / AR, or touch-screen-based interaction \cite{mulloni2010zooming, buschel2019investigating}. Lee et al. \cite{lee2012automatic} proposed an automatic control of the zoom level based on the distance between the user's view and the target object, increasing the zoom level as the AR camera approaches the tracking target, enabling users to have a closer look of the content with less tracking failure. While they linked the zoom control to the movement of the user's view, the users held the AR camera in their hands. In comparison, our work investigates how head movement, as one of the hands-free input methods, could be useful for controlling zooming and panning to explore large, high-resolution images in VR.

\subsection{Hands-Free Input Methods}
Advancements in human-computer interaction have led to the development of alternative input methods beyond traditional keyboard and mouse interfaces. Hand gesture tracking as input modality has been actively developed and used as a natural interaction method for 3D interaction in both VR~\cite{sagayam2017hand, yang2019gesture} and AR~\cite{satriadi2019augmented, billinghurst2014hands} environments. 
Despite its versatility and wide adoption in VR/AR interfaces, one of the typical limitations of hand gesture-based input is \textit{gorilla arm} syndrome, where prolonged arm use leads to fatigue \cite{gorillaarm_fatigue}.
As an alternative, hands-free input methods~\cite{piumsomboon2017gaze,handsfree_interfaces} have gained attention due to their potential to improve accessibility and usability. 
Hands-free interaction includes input modalities that do not require physical contact with a controller. Common approaches include voice commands, eye tracking, and head tracking \cite{handsfree_interfaces}.

\textbf{Speech Control.} Speech interfaces allow users to issue commands verbally or dictate text input \cite{clark2019state, lu2024human}. Speech interaction has been considered a natural and efficient interaction modality in XR systems \cite{muller1998speech, piumsomboon2014grasp}, especially in combination with gesture-based interaction \cite{bolt1980put, lee2013usability}. Although effective for discrete commands, speech is less suitable for continuous or highly granular adjustments \cite{hepperle20192d,igarashi01}.

\textbf{Eye Tracking.} Eye tracking technology has been actively researched in the context of human-computer interaction~\cite{majaranta2014eye, eyetracking_survey}. Gaze direction is a natural way to control an interface, as the eyes instinctively focus on areas of interest, and has also been actively investigated in VR \cite{piumsomboon2017gaze, pfeuffer2017gazepinch} and AR \cite{kyto2018pinpointing, park2008wearable} interfaces. Modern eye trackers using infrared corneal reflections achieve high accuracy \cite{majaranta2014eye, eyetracking_survey}. A key challenge, however, is the \textit{Midas touch problem}, which makes it difficult to distinguish intentional interactions from normal eye movements \cite{jacob1990midastouch, majaranta2014eye}. Although some modern VR/AR HMDs integrate eye-tracking technology, this technology is mostly supported on high-end devices. It has not reached the status of being widely adopted as a standard feature.

\textbf{Head-Based Interaction.} Head tracking has been studied to control computers using natural head movements \cite{hunke1994face, strumillo2012vision}. Early work in human-computer interaction identified head movement as a simple and effective pointing method, often more accurate than voice or brain-computer interfaces \cite{eyetracking_survey}. It is also considered to be more affordable compared to eye-tracking technology.
Head-based input is especially useful in assistive technology, allowing users with limited limb mobility to move a cursor or make selections by moving their head, using a webcam or infrared sensor.
Systems such as Camera Mouse \cite{betke2002camera, nabati2010camera} and Enable Viacam (eViacam)\footnote{\url{https://eviacam.crea-si.com}} track the face of a user through a webcam, translating head movements into pointer control \cite{headtracking_pathology}. These systems have been used successfully by individuals with high spinal cord injuries, as head and eye movements remain intact \cite{betke2002camera, eyetracking_survey}.
Harrison and Dey’s \textit{Lean and Zoom} system 
\cite{harrison08} detected users leaning toward or away from a desktop screen, automatically magnifying the content. Users reported that this head-driven zooming felt natural and fluid, making the text easier to read and reducing discomfort.

Head tracking is a key enabler technology for modern VR/AR \textcolor{blue}{head-mounted displays} (HMDs) to control the user’s viewpoint, inducing a sense of presence in VR \cite{slater2018immersion, wu2019effect}. \textcolor{blue}{In addition, head tracking has been also actively used as a hands-free interaction method, such as using the head direction for pointing \cite{atienza2016, qian2017, yan2020, monteiro2023}}. \textcolor{blue}{Apart from using head-based pointing for selection, Rudi et al. \cite{rudi2016} proposed using head-orientation based pointing to zoom and pan a map shown on a wearable display. Their method checked if the pointer went outside the set boundary to trigger zoom or pan.} \textcolor{blue}{Wang et al. \cite{wang2025} proposed and tested dynamic control-display gain to control head-orientation based pointer. Their method used horizontal and vertical head rotation to control the direction of pointing.} \textcolor{blue}{These prior work mainly used and manipulated the head orientation but did not investigate anteroposterior movements as in our work.}

\textcolor{blue}{Researchers also actively investigated head gestures as an input method. Hou et al. \cite{hou2023} proposed a classifier to distinguish head gestures from head pointing to utilize both as hands-free interaction methods in a VR system. Yan et al. \cite{yan2018} investigated various head gestures identified through an elicitation study. While they did use anteroposterior movements to control zoom as in our work, their approach was to recognize the head gesture to trigger zoom in or out, compared to our approach which continually maps the head movement to the zoom level.}

While prior work has thoroughly explored zooming and panning techniques across desktop, mobile, and AR/VR platforms, most approaches rely on hand-held controllers, touch interfaces, or indirect mappings such as pressure sensors or gesture recognition. These methods often impose physical constraints or require significant setup, making them less suitable for hands-free continuous interaction in immersive contexts. Existing hands-free alternatives—such as gaze-based or voice-controlled interfaces—struggle with precision or lack suitability for granular control. Although head tracking has been used for cursor control and viewpoint manipulation, its potential for fine-grained, dual-purpose navigation (zoom-and-pan) in immersive 2D content exploration remains underexplored. Moreover, previous systems typically rely on singular head-motion mappings (e.g., lean-to-zoom) and lack comparative evaluations of multiple head-based techniques within the same task framework. To address this gap, we introduce and evaluate \textit{HeadZoom}, systematically studying three distinct interaction techniques within a controlled experimental setting. 


\section{The HeadZoom Prototype}
\begin{figure}[htb]
    \centering
    \includegraphics[width=0.6\linewidth]{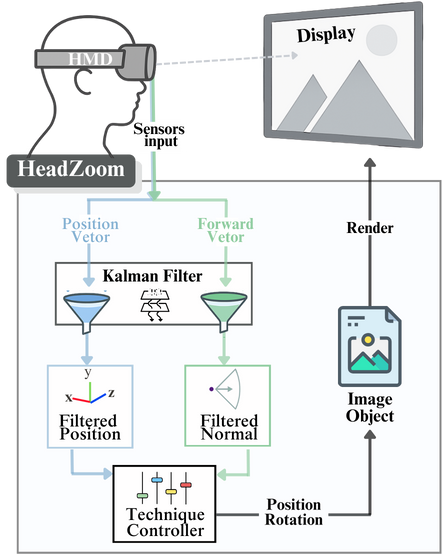}
    \caption{Block diagram of the HeadZoom system}
    \label{fig:headzoom_summary}
\end{figure}

The HeadZoom prototype was developed in Unity3D (v6) for Meta Quest 3 headsets. HeadZoom is a hands-free interaction framework that provides precise, controller-free zooming and panning in VR. As shown in \autoref{fig:headzoom_summary}, the system continuously captures the HMD position and orientation, smoothing through Kalman filters to suppress jitter while preserving responsiveness. A virtual image plane \textcolor{blue}{measuring 2 meters by 1 meter is dynamically placed at the user’s eye height and centred 2 meters in front of the user. The plane’s normal is aligned with the user’s horizontal forward direction (i.e., the forward vector with its vertical component set to zero), ensuring that the image plane remains stably in front of the user, regardless of head tilt.} 


The HMD’s \textit{position vector} provides real-time 3D coordinates of the user’s head in space. This vector is used to compute the user’s distance from the image plane, which modulates the zoom level; closer positions result in a zoom-in effect, while retreating causes a zoom-out. The \textit{forward vector} represents the HMD directional orientation, specifically the normalized vector pointing where the user is facing. This vector is crucial for raycasting: It projects a ray from the user’s current position toward the image plane. Raycasting from the filtered head pose determines the target location for interaction. This plane-based mapping enables natural, spatially consistent simultaneous panning and zooming using only head motion. The intersection of this ray with the plane determines the panning target, allowing the system to infer where in the image the person is attending to. Both vectors are filtered to reduce noise before being passed to the technique controller. \textcolor{blue}{Zoom and pan values are clamped to predefined minimum and maximum limits to prevent content from becoming inaccessible or unstable. If tracking loss or abnormal data (such as sudden outliers) is detected, the system retains the last valid position until reliable data resumes.}
\begin{figure*}[t]
    \centering
    \includegraphics[width=0.7\linewidth]{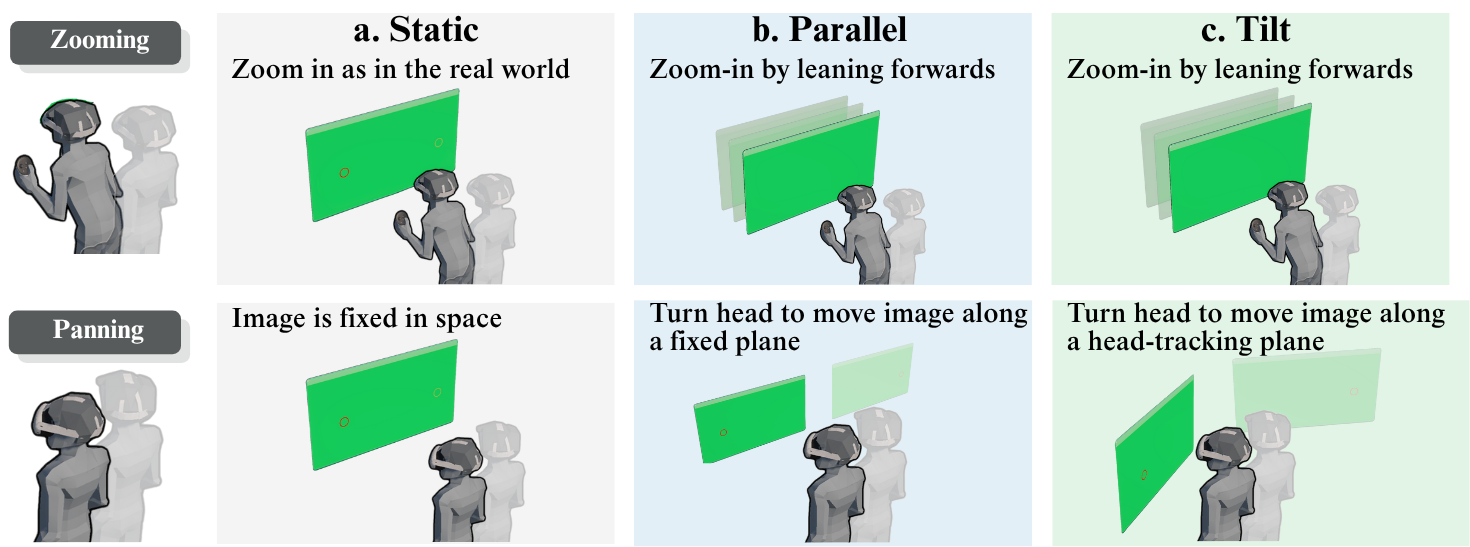}
    \caption{HeadZoom interaction modes:  (a) static, (b) parallel, (c) tilt}
    \label{fig:interactionScenarios}
\end{figure*}
\textcolor{blue}{\subsection{Calibration and Control-Display Mapping}}
\textcolor{blue}{To accommodate the natural diversity in users’ body sizes, postures, and ergonomic preferences, HeadZoom incorporates a brief, user-specific calibration phase. The primary goal of this calibration is to ensure that the mapping between physical head movement and virtual zoom is both intuitive and safe, maximizing the usable range of interaction without causing discomfort or requiring unnatural motion.
}
\textcolor{blue}{By capturing each user’s comfortable forward and backward lean limits, the system establishes personalized boundaries for zooming. This normalization process allows the full range of virtual zoom to be accessed within each individual’s natural movement envelope, eliminating the need for manual gain adjustment and preventing both overextension and underutilization. The virtual image is always positioned relative to the user’s initial pose, and as the user leans forward or backward, their current position is continuously normalized within their calibrated range. This ensures that, for every user, the closest and farthest zoom positions correspond precisely to their comfortable limits, with the image position clamped beyond these points to maintain stability and prevent accidental loss of content.}

\textcolor{blue}{This individualized, parameterless mapping guarantees that all subsequent interactions remain responsive, accessible, and ergonomically appropriate, regardless of user variability. By prioritizing user comfort and control, the calibration phase underpins the system’s ability to deliver precise, hands-free navigation for a wide range of participants.}

\textcolor{blue}{\subsection{Interaction Techniques}}
As illustrated in Figure~\ref{fig:interactionScenarios}, HeadZoom supports three head-driven interaction techniques, 
\textcolor{blue}{each designed to systematically explore different mappings between head motion and image navigation, with an emphasis on spatial consistency, geometric predictability, and user control.}


\textcolor{blue}{In the \textit{Static} condition, the image remains completely stationary throughout the interaction: neither zooming nor panning is supported. All movements or rotations of the user's head do not affect the position, size, or orientation of the image. This mode serves as a passive baseline for comparison, isolating the effects of head-driven navigation.}


\textcolor{blue}{The \textit{Parallel} condition implements a direct geometric mapping between head translation along the Z axis (forward/backward) and zoom level, using a linear function based on the user’s calibrated range. Panning is realized by raycasting from the user’s head position in the view direction onto a static image plane. This mapping ensures that as the user approaches the image, the angular range for lateral (yaw) and vertical (pitch) panning increases, providing broader coverage at close distances and finer control at greater distances. At the neutral position (\~2~m from the image), the maximum attainable yaw is $\pm 45^\circ$ and pitch is $\pm 26.6^\circ$; these values increase as the user leans forward. The image orientation remains fixed, which supports spatial predictability and minimizes disorientation during navigation. This approach leverages the geometric relationship between head position and the projected field of view, enabling efficient exploration of large images.}



\textcolor{blue}{The \textit{Tilt} condition extends Parallel by dynamically coupling the image plane’s orientation to the user’s head pose. The image rotates in yaw and pitch to always face the user, and in roll to match the user’s head tilt. The zoom mapping remains the same as Parallel. In this mode, the maximum attainable yaw and pitch at the neutral position are $\pm 36.9^\circ$ and $\pm 20.6^\circ$, respectively, with these angles increasing as the user moves closer. Unlike Parallel, Roll is not artificially limited; the image rotates exactly with the user’s head, preserving alignment and supporting a consistent spatial relationship between the user and the content. This design reduces rotational distortion and maintains geometric coherence during navigation.}

\textcolor{blue}{\textit{Note:} In both Parallel and Tilt, the zoom effect is implemented as a dolly-in (camera movement) rather than a change in field of view, preserving perspective and spatial cues critical for visual search tasks.}\\

\subsection{Filtering}
\textcolor{blue}{
Because neck and head muscles naturally exhibit continuous micro-movements to maintain balance, our initial static Kalman filters—manually tuned via trial and error—smoothed most jitter but proved insufficient at higher zoom levels, where the motion-amplifying mapping magnifies even tiny shakes. To compensate, we made each filter’s Process Noise Covariance (Q) and Measurement Noise Covariance (R) dynamic functions of the user’s normalized head position along the calibrated lean interval. Specifically, we define $X\in[0,1]$ so that zero maps to the maximum backwards lean, 0.5 to the neutral pose, and 1 to the maximum forward lean; Y is the filter parameter value. At runtime, the headset’s current X-value selects Q and R by sampling from two piecewise curves inscribed on the X–Y plane.}

\textcolor{blue}{
For the Tilt interaction, we hold Q constant at Y=0.01 across the entire range ($X\in[0,1]$). R remains flat at Y=0.0001 for backward-to-neutral positions ($X \in [0,0.5]$), then increases linearly from 0.0001 up to 0.1 as X goes from 0.5 to 1. In contrast, the Parallel interaction uses a mirrored Q curve—flat at 0.01 between X=0 and 0.5, then decreasing linearly to 0.0001 between X=0.5 and 1—while R follows the same piecewise profile as in Tilt. This position-based filtering adaptively dampens jitter at all zoom levels, eliminating image jerks without introducing perceptible lag.}

\vspace{0.2cm}
\section{Experiment}
We conducted a controlled user study to evaluate the usability and effectiveness of our HeadZoom interaction technique in a digital photography scenario. The study aimed to assess participants' performance and subjective experience in different zooming methods in Virtual Reality (VR) environments. We collected both quantitative measures (task completion time, success rate, head movement) and qualitative feedback to understand how each technique affects user engagement, task efficiency, and overall comfort.


\begin{figure*}
    \centering
    \includegraphics[width=0.85\linewidth]{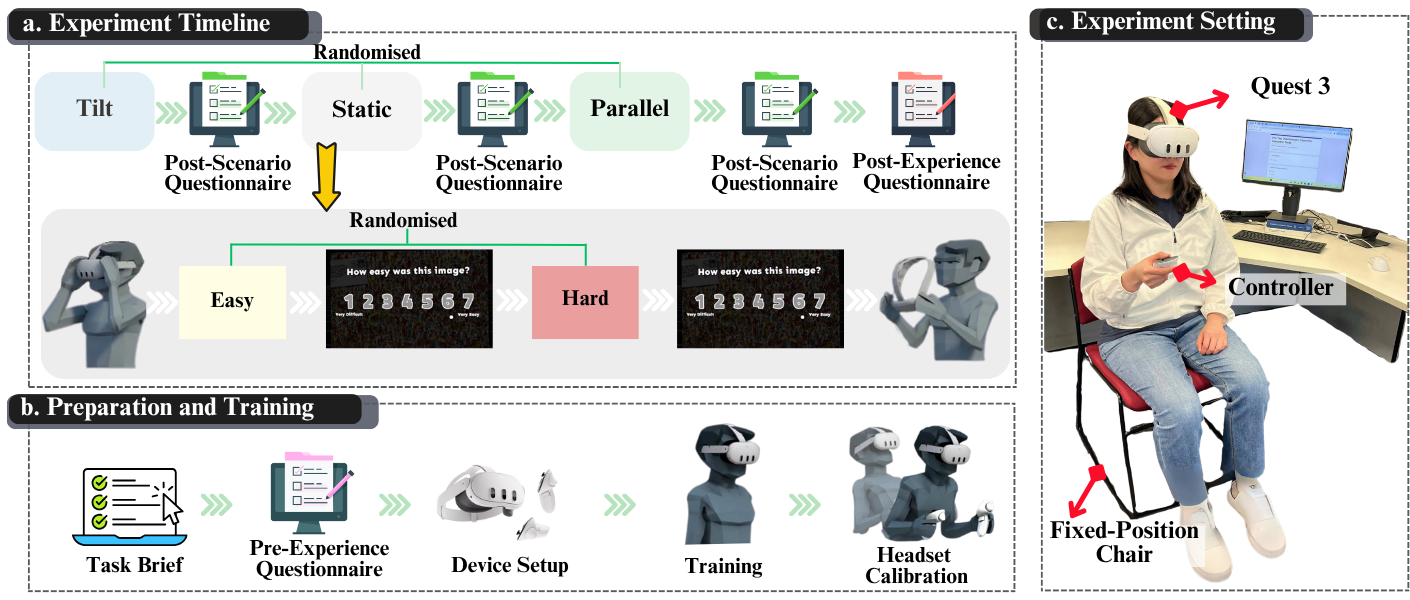}
    \caption{ Experiment study design: (a) Experimental Timeline, (b) Preparation and Training process, (c) Experiment Setting with one participant.}
    \label{fig:experiment}
\end{figure*}

\subsection{Task}
To reflect the panning and zooming demands in digital photography scenarios, we adopted \textit{Where’s Wally} \cite{Wally} task, which was also used in Brown et al. \cite{TaskChoose}’s 2D visual search study. \textit{Where’s Wally} is a well-known visual search activity that is easy to explain but challenging to execute, as its large images require substantial panning and zooming. Moreover, this task provides images with varying difficulty levels, enabling us to select functions of different complexities to comprehensively evaluate our Head Zoom interaction technique.

We selected seven images from \textit{Where's Wally}.  An image, featuring the fewest characters and minimal distractions, was used for training so that the participants could familiarize themselves with the three interaction modes (Static, Parallel, and Tilt). The remaining six images were classified into \textcolor{blue}{two difficulty levels (three easy and three hard)} according to difficulty based on approximate character count and background complexity.

\subsection{Task Image Characteristics and Visual Challenge}
The pictures selected for the experiment 
came from the "Where's Wally?"\footnote{Where’s Wally? Images © Martin Handford / Walker Books Ltd. Used under fair use for academic research purposes. All images used in the study, including training and trial stimuli, are available as supplementary materials to this article.} series and varied considerably in visual complexity. 

Each image depicts a complex, cartoon-style environment populated by dozens to hundreds of small, visually similar characters, distributed across spatially and semantically varied backgrounds. Despite their playful appearance, these scenes pose significant cognitive and perceptual challenges for visual search tasks.

The training image, \textbf{city}, is the least complex of the set, featuring a modest number of characters with relatively clear segmentation and minimal foreground clutter. In contrast, the remaining six images vary widely in layout and visual load. Others, e.g. \textbf{beach}, introduce dense and unstructured clustering of figures, many partially occluded by umbrellas and beach props, creating substantial camouflage effects. Other images are highly populated but exhibit more regular spatial organization.

The most visually demanding scenes combine extreme crowd density with intricate environmental elements, such as market stalls, fortress walls, snowy slopes, intense color variation, and overlapping figures. These images often present large contiguous areas of visual noise, making it difficult to distinguish targets based solely on spatial cues. The combination of character repetition, distractor similarity, and environmental detail results in high visual entropy, amplifying the difficulty of head-driven zoom and pan interaction.

This deliberate diversity in complexity ensured that the interaction techniques were evaluated across a realistic spectrum of search conditions, ranging from moderately challenging to perceptually saturated scenes.

\subsection{Task Prototype Design}
We developed our task prototype in Unity3D (version 6) for the Meta Quest 3 VR HMD. As shown in \autoref{fig:teaser}.c, each trial presented participants with a large Where’s Wally image (2800×1749 pixels). A red ring (approximately 105 pixels in radius) overlaid the image to indicate the current raycast hit point. In the upper left corner of the image, the participants saw the assigned interaction mode (e.g. static, parallel, or tilt) and the trial number, ensuring that they were aware of the condition they were experiencing.

Before the first trial, participants calibrated the headset to establish how far forward and backward they could tilt their heads for maximum zoom‐in and zoom‐out. Their neutral head position was recorded as the central reference point. During the trial, participants used the designated interaction technique, moving their heads following these calibrated lean thresholds to zoom in or out. Panning occurred naturally through head rotation or subtle adjustments in body posture. Once Wally was centred in the red ring, participants pressed the controller’s trigger to indicate they had located the target.

Each trial allocated participants two minutes to locate Wally and allowed up to three attempts. A trial ended if participants successfully aligned Wally within the red ring, failed three times, or exceeded the two-minute limit. 

All trial-related data, including video recordings, sensor readings, and image metrics, are logged in tab-separated CSV files. We also captured participants' head orientation (pitch, yaw, roll) throughout each attempt for further analysis.


\subsection{Experimental Design}\label{subsec:Experiment}
Our study adopted a within subjects design for the participants that included three interaction modes (Static, parallel, and tilt) and two difficulty levels (easy \textcolor{blue}{ and hard}). \autoref{fig:experiment}.a shows an overview of the experimental timeline. Each participant was randomly assigned one low and \textcolor{blue}{one hard} difficulty image under each mode, resulting in six total trials (3 modes x \textcolor{blue}{2} difficulty levels). We counterbalanced and randomized the order of both mode and difficulty levels to mitigate the potential bias from time‐sequence effects or subjective judgments.

Each condition consisted of two trials (one low-difficulty image and one \textcolor{blue}{high-difficulty} image), presented in a counterbalanced sequence. Upon finishing each trial, participants completed a subjective rating (on a 1–7 scale) to indicate their perceived challenge in locating Wally within that image.

After completing both trials in a scenario, participants removed the headset and filled a post-condition questionnaire assessing system usability and user experience. Upon concluding all three modes (i.e., six trials), they completed a final post-experience questionnaire, ranking each interaction technique in terms of zooming efficiency and object-finding accuracy. 

\subsection{Procedure}
\autoref{fig:experiment}.b illustrates the overall experimental setup. Upon arrival, participants were given an information sheet describing the procedures, followed by completion of a consent form and a pre‐experience questionnaire. Then, we configured the Quest headset and seated participants in a fixed‐position chair (\autoref{fig:experiment}.c). We used Meta Horizon to stream the headset’s view for monitoring participant performance. Participants then underwent a 15 minute training session, familiarising themselves with all three interaction modes for the \textit{Where's Wally} task. Participants then proceeded according to the process depicted in \ref{subsec:Experiment}. We used Google Forms to collect all pre-, post‐condition, and post‐experience responses. The experiment lasted approximately 40 minutes, after which each participant received a \$20 gift card as compensation.

\subsection{Participants}
We recruited 31 participants (16 women, 15 men, ages 19 to 49). A total of 8 participants (25.8\%) reported never using VR, 19 participants (61.3\%) had fewer than 5 hours of VR experience, 1 (3.2\%) reported having 5–20 hours of experience, 8 participants (25.8\%) had logged more than 20 hours, and 3 participants (9.7\%) had never used VR. Among all participants, 8 (25.8\%) 
had never experienced VR‐related motion sickness, 4 (12.9\%) were unsure or had insufficient VR experience,  and 19 (61.3\%) experienced it only occasionally.

\section{Metrics}
To assess user performance and understand the impact of each interaction technique, we computed quantitative metrics from the headset data logs, grouped into four categories: \textit{Time Completion}, \textit{Spatial Metrics}, \textit{Error‐Related Metrics}, and \textit{Subjective Difficulty}. 

\subsection{Time Completion Metrics}

This category evaluates how quickly and successfully participants locate Wally. The main metric is \textbf{Completion Time}, which measures the time taken to correctly identify Wally.



\subsection{Spatial Metrics}

These metrics quantify physical interactions and exploration patterns, including head movement, rotation, lean, zoom usage, and participant preferences.

\begin{itemize}[topsep=30pt,nosep,label={},leftmargin=0pt]
\item\textbf{Total Head Movement.} It is the cumulative headset displacement where \(\mathbf{p}_i\) represents headset position at frame \(i\).: $TotalHeadMovement =  \sum_{i=1}^{N-1} |\mathbf{p}_{i+1} - \mathbf{p}_i$.

\item\textbf{Total Head Rotation.} Summatin of angular displacement between consecutive forward vectors \(\mathbf{f}_i\): $\text{TotalHeadRotation} = \sum_{i=1}^{N-1} \arccos\left(\frac{\mathbf{f}_i \cdot \mathbf{f}_{i+1}}{\|\mathbf{f}_i\|\|\mathbf{f}_{i+1}\|}\right)$.
\item\textbf{Lean Usage.} Measures maximum lean from the initial headset position across all frames.
\item\textbf{Hover Time on Characters.} Time spent, $\Delta t_i$ gazing near a specific character (Wally, Wenda, Wizard, Odlaw) $
\text{HoverTime}_{char} = \sum_{i=1}^{N} \text{isHovering}_{i,char} \times \Delta t_i$ 
with binary indicator $\text{isHovering}_{i,char} = [\|\mathbf{g}_i - \mathbf{h}_{char}\| < \text{HitRadius}$.
\item\textbf{Zoom Usage.} Evaluates how participants interact with the zoom functionality. These metrics help characterize participants' zoom interaction behaviors, reflecting how actively they used zoom capabilities. They are measured through different metrics:
\item\textbf{Zoom Change Count} counts how many times participants meaningfully change the zoom level during the trial. 
\item\textbf{Total Zoom Distance} corresponds to the cumulative amount participants zoomed in and out across the entire trial.
\item\textbf{Average Zoom} indicates participants' average zoom level throughout the trial. Similar definitions apply for \textbf{AvgHeadRotation} and \textbf{AvgHeadMovement}.  \textbf{Maximum Zoom} records the highest level of magnification achieved during the trial.  
\end{itemize}
\subsection{Error‐Related Metrics}

These metrics reflect errors made during task execution:

\noindent
\textbf{False Positives.} Count of incorrect identification attempts.

\noindent
\textbf{Task Success Rate.} A binary metric indicating task success (finding Wally under 120 seconds and within three tries).

\subsection{Subjective Difficulty Metrics}
Participants rated each image's difficulty using a 1--7 scale (\texttt{ImageUserGrade}), where 1 = ``very difficult" and 7 = ``very easy". This allows correlation analysis with objective performance metrics (e.g., Completion Time, False Positives) showing discrepancies or consistencies between perceived difficulty and actual performance.

\section{Results}
\label{sec:results}
\begin{figure}[!h]
\resizebox{\columnwidth}{!}{
	\includegraphics{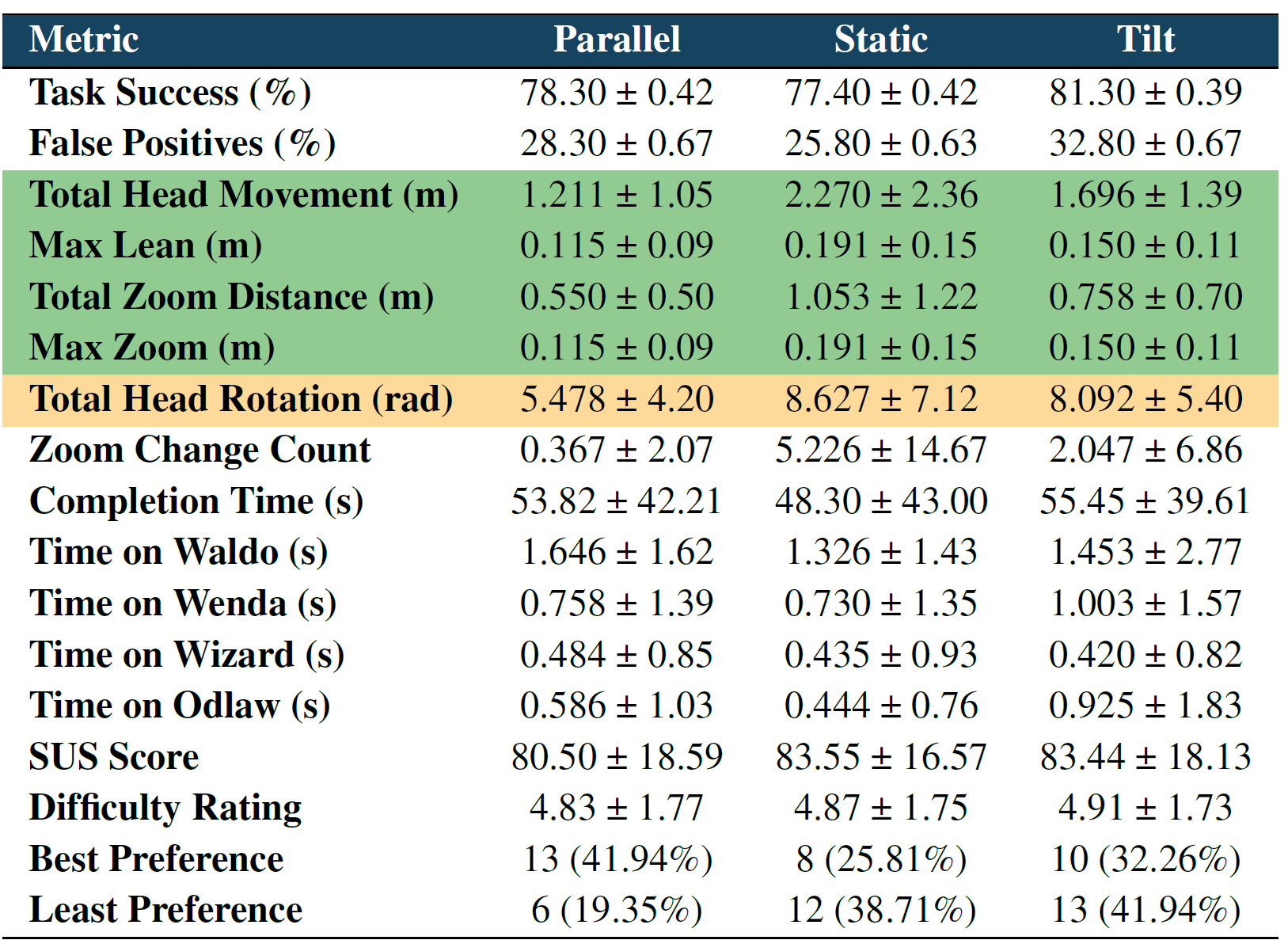}
}
\caption{Summary of task performance and behavioral metrics by interaction mode (mean $\pm$ SD). Green rows indicate high significance ($p \leq 0.01$), yellow rows moderate significance ($0.01 > p \leq 0.05$), and uncolored rows are not significant ($p > 0.05$).}
\label{tab:sum_interaction_metrics}
\end{figure}

To evaluate the performance and interaction characteristics of the three interaction techniques (Static, Parallel, and Tilt), we performed a comprehensive statistical analysis on data collected from a controlled VR experiment involving 31 participants. The task involved participants in locating a target character (Wally) within images of varying complexity. The statistical evaluation process consisted of three main stages and is presented in Figure \ref{tab:sum_interaction_metrics}.

\vspace{0.2cm}
\noindent
     \textbf{Repeated Measures Analysis of Variance (ANOVA)} was used to identify general differences between the three interaction techniques for each metric. We used repeated measures ANOVA rather than one-way ANOVA because each participant experienced all three interaction conditions (Static, Parallel, and Tilt), creating within-subject dependencies in the data. This approach accounts for individual participant differences and provides greater statistical power by reducing error variance. We considered the differences statistically significant at a $p$ value $\leq 0.05$.

\vspace{0.2cm}
\noindent
\textbf{Post-hoc pairwise comparisons} were made using paired t-tests to pinpoint specific differences between interaction techniques for metrics identified as significant by ANOVA. To address the issue of multiple comparisons that inflate Type I errors, we applied the Bonferroni correction, adjusting the significance threshold to $\alpha = 0.0167$ ($0.05$ divided by $3$ comparisons per metric).

\vspace{0.2cm}
\noindent
    \textbf{Effect sizes} were calculated using Cohen's d to quantify the significant differences. Cohen's d was calculated and used to compute Effect sizes. We interpret these as negligible $( d < 0.2 )$, small $(0.2 \leq d < 0.5)$, medium $(0.5 \leq d < 0.8)$, or large $(d \geq 0.8)$.

\subsection{Interaction Pattern Differences}

The ANOVA results revealed \textit{significant differences} among interaction modes for $10$ out of $18$ metrics analyzed\textcolor{blue}{, with F-statistics ranging from $F(2,183) = 3.54$ to $F(2,183) = 63.00$ (all $p < 0.05$). Effect sizes ranged from $\eta^2 = 0.037$ to $\eta^2 = 0.408$}. The most pronounced differences were observed for Average Head Rotation (\textcolor{blue}{$F(2,183) = 63.00$, $p < 0.001$, $\eta^2 = 0.408$}) and Average Head Movement (\textcolor{blue}{$F(2,183) = 34.23$, $p < 0.001$, $\eta^2 = 0.272$}). 
After the Bonferroni correction \textcolor{blue}{($\alpha = 0.0167$)}, \textcolor{blue}{$15$ out of $54$} pairwise comparisons remained statistically significant\textcolor{blue}{. Post-hoc $t$-statistics ranged from $t(84.69) = 0.036$ to $t(88.14) = 10.385$, with degrees of freedom varying from $84.69$ to $124.00$ depending on equality of variances, and Cohen's $d$ effect sizes ranging from $d = 0.006$ to $d = 1.871$}. The key differences were:

\vspace{0.2cm}
\noindent
\textbf{Static Mode} required \textit{significantly more head movement} ($M = 2.27$, $SD = 2.34$) and rotation ($M = 8.63$, $SD = 7.06$) compared to Parallel mode (both $p < 0.01$). Static mode also showed \textit{significantly higher average lean} ($M = 0.09, SD = 0.10$) and maximum zoom ($M = 0.19, SD = 0.15$), indicating greater user physical effort.

\vspace{0.2cm}
\noindent
\textbf{Parallel Mode} exhibited the least head movement ($M = 1.21, SD = 1.04$) and rotation ($M = 5.48, SD = 4.17$), with \textit{significantly fewer zoom changes} ($M = 0.37, SD = 2.05$), indicating a more constrained and efficient interaction pattern.

\begin{figure}[b!]
    \centering
    \includegraphics[width=0.7\columnwidth]{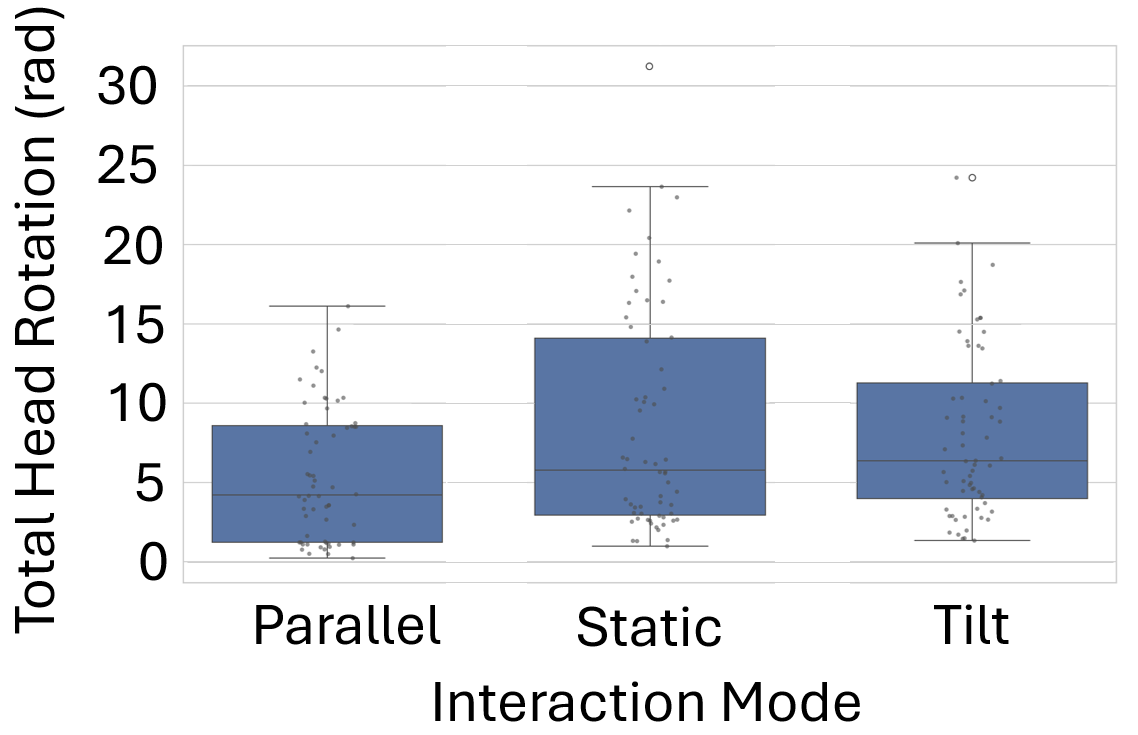}
    \caption{Total head rotation (radians) across the three interaction modes. Static mode resulted in the greatest head rotation, indicating higher physical effort, whereas Parallel mode showed the least, reflecting a more \textcolor{blue}{physically} constrained interaction.}
    \label{fig:rotation_metrics}
\end{figure}

\begin{figure}[t!]
    \centering
    \includegraphics[width=0.7\columnwidth]{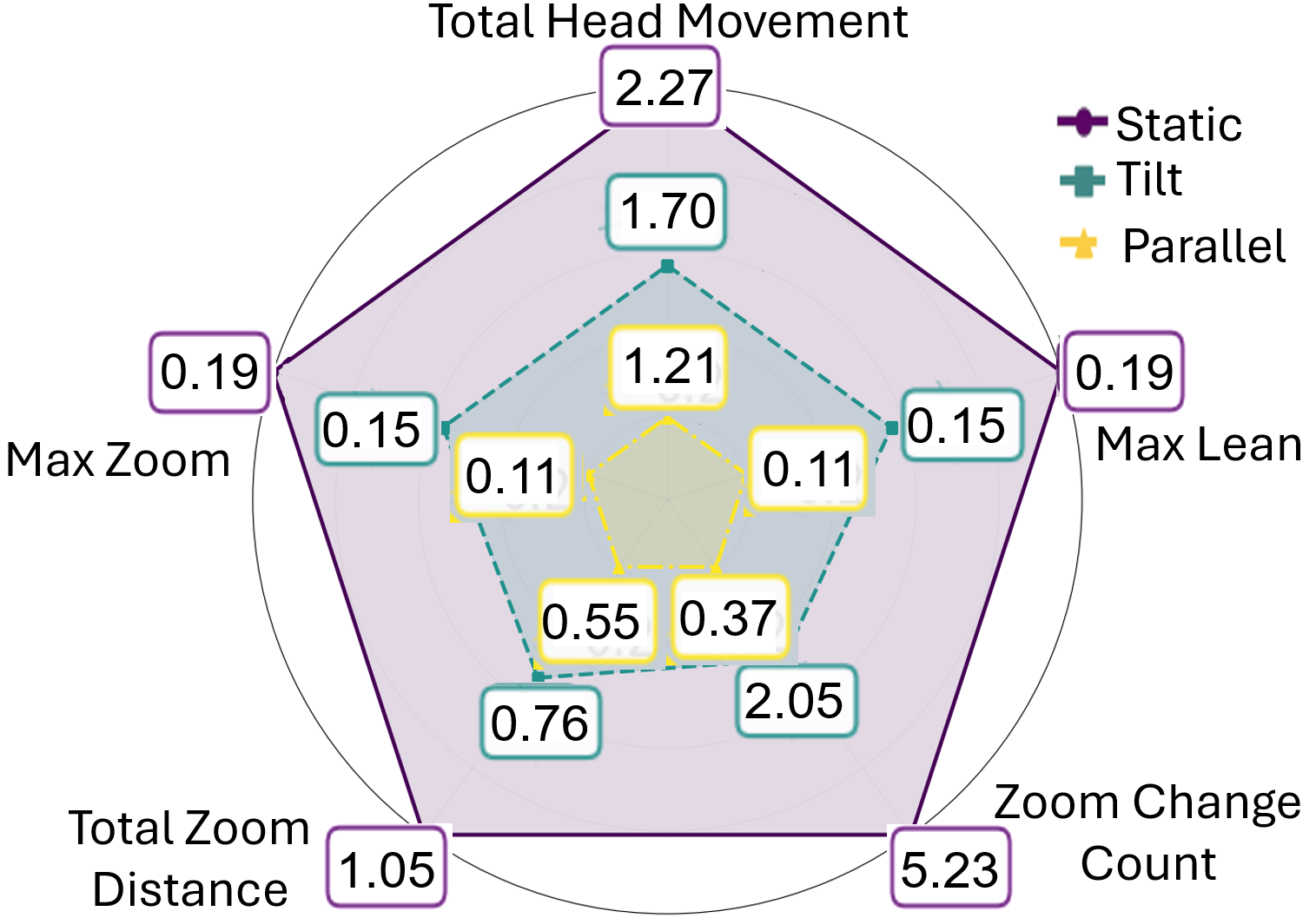}
    \caption{Comparison of normalized average performance metrics across interaction modes. Metrics are the Total Head Movement, Max Zoom, Total Zoom Distance, and Zoom Change Count. Lower values indicate better performance outcomes and \textcolor{blue}{less physical demand}.}
    \label{fig:radar_metrics}
\end{figure}

\vspace{0.2cm}
\noindent
\textbf{Tilt Mode} generally showed intermediate values for most physical metrics, indicating moderate physical interaction compared to the other two modes. \autoref{fig:radar_metrics} shows the normalized averaged performance metrics for Total Head Movement, Max Zoom, Total Zoom Distance, Zoom Change Count and Max Lean, revealing that Parallel is the \textcolor{blue}{physically less demanding} mode, as it leads to lower physical movements. \autoref{fig:rotation_metrics} presents the distributions of the Total Head Rotation per interaction mode, indicating Static mode required more movements when compared to the other interaction modes.

\vspace{0.2cm}
\noindent
\textbf{Effect sizes} further clarify the magnitude of these differences, highlighting the large effect sizes between Static and Parallel for AvgHeadRotation (\textcolor{blue}{$d=1.87$}) and AvgHeadMovement (\textcolor{blue}{$d=1.32$}). Similarly, we observed large effect sizes between Parallel and Tilt for AvgHeadRotation (\textcolor{blue}{$d = 1.42$}) and between Static and Tilt modes for AvgHeadRotation (\textcolor{blue}{$d=0.88$}) and AvgHeadMovement (\textcolor{blue}{$d=0.85$}).

\subsection{Task Performance and Efficiency}

\begin{figure}[t!]
    \centering
    \includegraphics[width=0.8\columnwidth]{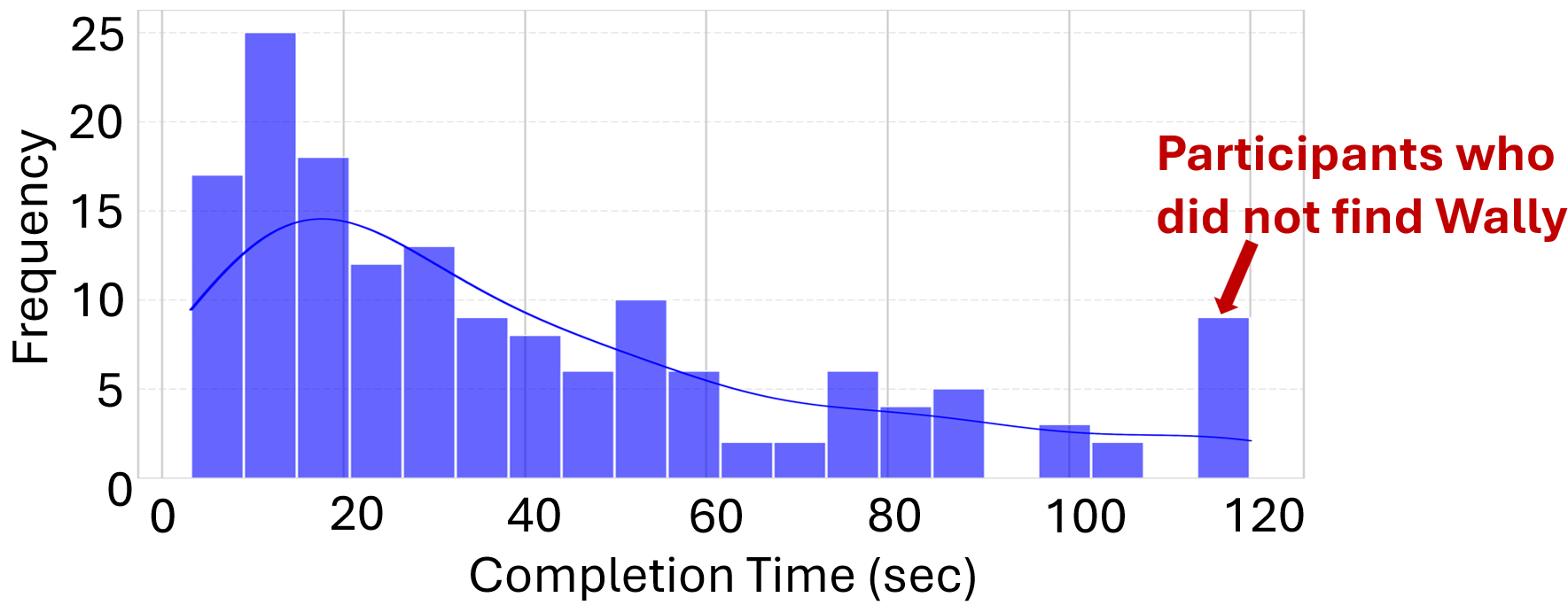}
    \caption{Distribution of task completion times (in seconds) across all interaction modes and participants. Most tasks were completed within 60 seconds, indicating generally efficient performance, though notable variability exists among participants. The rightmost task represents all participants who could not find Wally (completion time = 120s)}
    \label{fig:time_completion}
\end{figure}

Despite significant differences in physical interaction patterns, \textit{no statistically significant differences emerged} regarding task completion time (\textcolor{blue}{$F(2,183) = 0.51$, $p = 0.604$}). Mean completion times were similar across modes: Static with $48.30$ secs ($SD = 42.65$), Parallel with $53.82$ seconds ($SD = 41.85$), and Tilt with $55.45$ seconds ($SD = 39.30$). Additionally, \textit{no significant differences were found} for \textit{false positive rates} \textcolor{blue}{($F(2,183) = 0.61$, $p = 0.544$)} or \textit{subjective difficulty ratings} \textcolor{blue}{($F(2,183) = 0.03$, $p = 0.970$)}. This suggests that all three interaction techniques provided \textit{similar efficiency and accuracy}, despite differences in physical interaction demands. \autoref{fig:time_completion} shows the distribution of the completion times across participants.

\subsection{Usability Evaluation}
The System Usability Scale scores, assessing subjective usability, \textit{did not show significant differences between interaction modes} (\textcolor{blue}{$F(2,183) = 0.57, p = 0.568$}). All three modes achieved high usability scores: Static $83.55$ ($SD = 16.57$), Parallel $80.50$ ($SD = 18.59$), and Tilt $83.44$ ($SD = 18.13$).
These scores indicate excellent usability for all interaction methods. The analysis of individual SUS items also revealed that there were no significant differences in perceived complexity, ease of use, learnability, confidence, or support needs, confirming that differences in physical interaction did not significantly influence overall usability perceptions.

\subsection{User Preferences}
The Parallel mode was most frequently selected as the preferred method, with the highest ``best" ratings \textcolor{blue}{(13 participants, 41.94\%)}, while Static and Tilt modes were more frequently identified as the least preferred. Specifically, Parallel was chosen most frequently as the ``best" interaction mode by participants and less often marked as the ``worst" \textcolor{blue}{(6 participants, 19.35\%)}, contrasting with Tilt, which received the highest number of ``worst" ratings \textcolor{blue}{(13 participants, 41.94\%)}. \textcolor{blue}{Static mode received intermediate preference ratings with 8 participants (25.81\%) rating it as best and 12 participants (38.71\%) rating it as least preferred.} This subjective preference aligns with objective findings on reduced physical demand for the Parallel mode, underscoring user appreciation for its efficiency.

\subsection{Impact of Image Complexity}
\begin{figure}[b!]
    \centering
    \includegraphics[width=0.8\columnwidth]{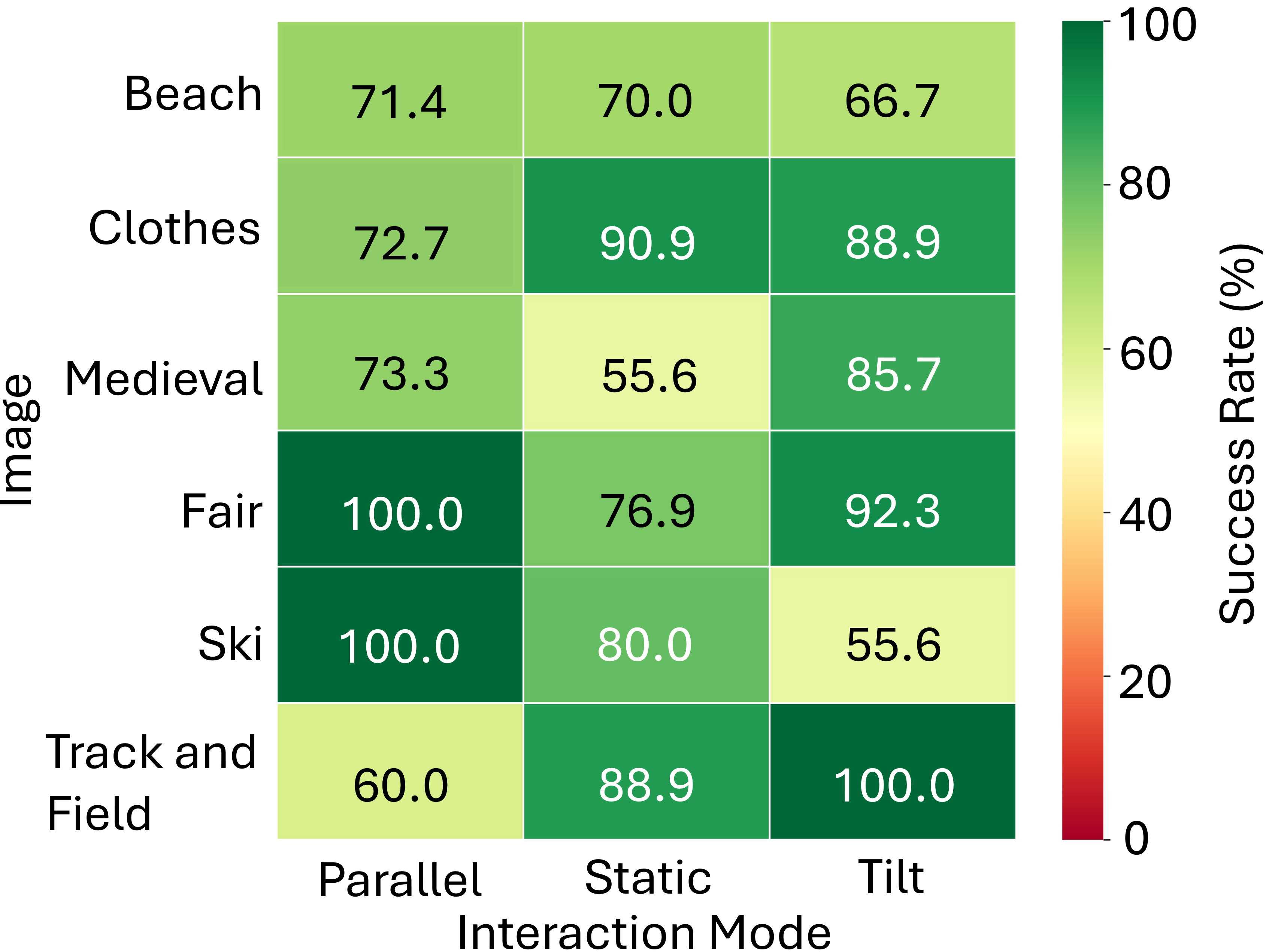}
    \caption{Success rates (\%) by interaction mode and image type. Darker green shades correspond to higher success rates and red to lower, emphasizing variability in performance across interaction methods and image visual complexity.}
    \label{fig:success_rates}
\end{figure}

``Time On'' metrics (TimeOnWally, TimeOnWenda, TimeOnWizard, TimeOnOdlaw) indicate the duration participants focused on specific targets or distractors did not differ significantly across interaction modes\textcolor{blue}{. Specifically, TimeOnWaldo ($F(2,183) = 0.38, p = 0.684$), TimeOnWenda ($F(2,183) = 0.68, p = 0.507$), TimeOnWizard ($F(2,183) = 0.17, p = 0.842$), and TimeOnOdlaw ($F(2,183) = 2.29, p = 0.104$) all showed non-significant effects}. \textcolor{blue}{However, \autoref{fig:success_rates} reveals substantial variability in success rates across images and interaction modes, with some images showing large differences (e.g., Medieval: $55.6\%$ to $85.7\%$, Ski: $55.6\%$ to $100.0\%$). This reflects inherent differences in difficulty in Where's Waldo images, where visual complexity and target detectability vary systematically across scenes.} This finding held across varying image complexities, suggesting consistent interaction patterns regardless of visual clutter.

\subsection{Gender Differences}
In terms of gender, only the User Difficulty Rating showed \textit{significant gender differences}, with males rating tasks as slightly easier ($M = 5.20$) compared to women ($M = 4.56$\textcolor{blue}{; $t(174.34) = 2.56, p = 0.011, d = 0.37$}). Further analysis revealed \textit{a significant gender-interaction mode effect for SUS scores specifically in the Parallel mode}, where males reported \textit{significantly higher usability scores} ($M = 86.00$) compared to females ($M = 75.00$\textcolor{blue}{; $t(50.88) = 2.36, p = 0.022, d = 0.61$}), indicating gender may influence subjective experience particularly in this interaction condition.

\section{Discussion}
\label{sec:discussion}
The following section discusses the implications and significance of our main experimental findings. Each finding is contextualized within the broader scope of VR interaction research, highlighting its relevance and potential impact on VR interface design and usability.

\vspace{0.2cm}
\noindent
\textbf{Finding 1: Significant Differences in Physical Interaction Patterns.} Our study revealed significant differences in physical interaction patterns among the three modes (Static, Parallel, and Tilt), particularly regarding head movement, rotation, and zoom usage. The Static mode had significantly higher physical engagement, evidenced by greater head movement and rotation, than the Parallel mode. Conversely, the Parallel mode demonstrated minimal physical exertion requirements, making it the most \textcolor{blue}{physically less demanding}. This aligns with the participants' subjective preferences, as the Parallel mode was frequently the preferred interaction mode and less frequently the least preferred. The Tilt mode generally showed moderate values, indicating that this technique balances physical engagement and comfort. This finding is important, as it suggests that interaction techniques significantly affect user comfort and could influence user adoption, especially in prolonged VR applications or precision-demanding tasks like medical imaging analysis.

\vspace{0.2cm}
\noindent
\textbf{Finding 2: Comparable Task Performance Across Interaction Modes.} Despite pronounced physical differences among interaction techniques, no significant differences were observed in task completion times, accuracy, or subjective difficulty ratings. This indicates that while the techniques differ significantly in \textcolor{blue}{physical demand}, they do not affect the effectiveness or accuracy of task execution. These results suggest that designers of VR interfaces have flexibility in choosing interaction techniques based on considerations other than raw performance metrics, such as user comfort, accessibility, or specific application requirements. Thus, the decision regarding interaction technique selection can prioritize user comfort and reduce physical strain without compromising task performance.

\vspace{0.2cm}
\noindent
\textbf{Finding 3: High Usability Scores Independent of the Interaction Technique.} All interaction modes received high usability ratings, with no significant differences observed in the overall SUS scores or in individual usability dimensions such as complexity, ease of use, learnability and confidence. This suggests that despite differences in how users physically interacted in VR, all methods provided an excellent user experience. High usability scores across all modes imply that users readily adapted to different interaction paradigms, reinforcing that usability perceptions are robust to variations in physical interaction patterns. Consequently, one can confidently employ varied interaction methods tailored to specific contextual considerations without concern for negative usability impacts.

\vspace{0.2cm}
\noindent
\textbf{Finding 4: Gender Differences in Interaction Experiences.} Our analysis identified gender-related differences, specifically in subjective task difficulty ratings and usability perceptions within the Parallel interaction mode. Males rated the tasks as slightly easier and reported significantly higher usability for the Parallel mode than females. These gender differences highlight the importance of considering demographic factors in interaction design, suggesting that user groups might perceive certain interaction methods differently. 

\vspace{0.2cm}
\noindent
\textbf{Finding 5: Robustness of Observed Differences.} Applying the Bonferroni correction minimally impacted the significant differences observed, with only a minor reduction from 16 to 15 significant pairwise comparisons. This highlights the robustness of our findings, indicating that the reported differences in physical interaction patterns and \textcolor{blue}{physical demand} are reliable and unlikely to be Type I errors. The large effect sizes observed, particularly between Static and Parallel modes, further substantiate the practical significance of these differences. These results reinforce the recommendation of the Parallel mode choice for extended or precision-based VR tasks.

\vspace{0.2cm}
\noindent
Our findings underscore the importance of \textcolor{blue}{physical demand} considerations in VR interaction design, demonstrate flexibility in achieving comparable performance outcomes across interaction methods, and highlight user preferences and demographic differences as critical considerations for user-centered VR system development.

\section{Application Scenarios}
Although the primary evaluation of HeadZoom focused on a controlled image search task, the system was designed with broader use cases in mind. To demonstrate the system’s generalizability, we explored how HeadZoom could be applied across several domains requiring precise, hands-free navigation of two-dimensional content.

\begin{figure}[t!]
    \centering
    \includegraphics[width=1\linewidth]{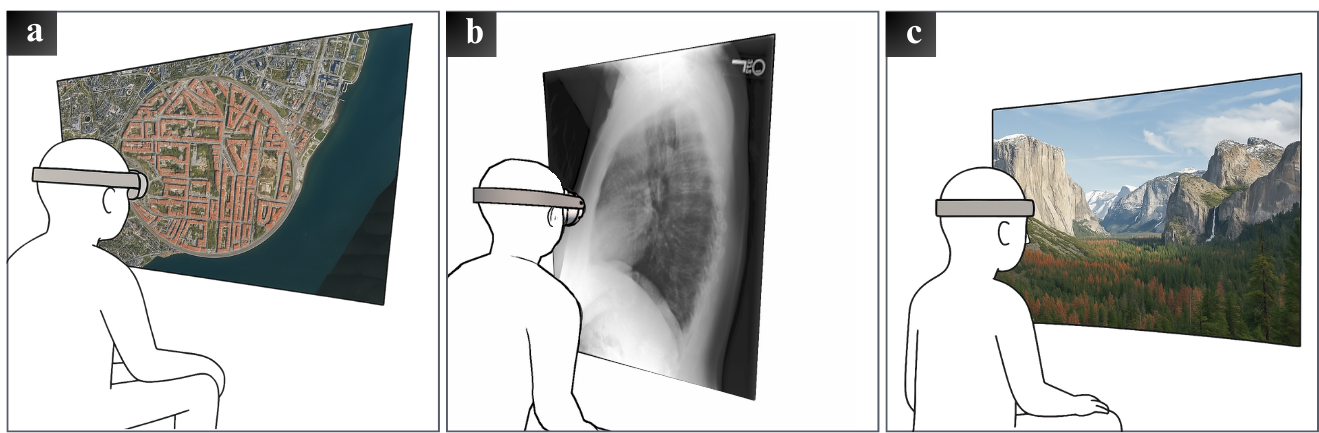}
    \caption{\textcolor{blue}{Headzoom interaction technology applied across three domains}: (a) Geographic Information System. (b) Medical imaging scenario showing radiological analysis of chest X-rays, and (c) Digital photography browsing via head-based navigation. }
    \label{fig:application}
\end{figure}

\textcolor{blue}{\textbf{Geographic Information System (GIS):} GIS interfaces rely on frequent panning and zooming using mouse-based interaction (desktop), or multitouch gestures (mobile) \cite{usability_zoom_pan,correia2005instory}. With HeadZoom, users can manipulate tiled map data by leaning forward and backward to zoom in or out, and by rotating their head to pan horizontally or vertically.  Users are tasked with locating iconic landmarks in globally recognized cities using only head-based input: leaning forward and backward adjusts the zoom level, while head rotations pan the map horizontally and vertically (see \autoref{fig:application}.a). We adopted a ‘focus zoom’ approach inspired by focus-plus-context methods \cite{baudisch_focuspluscontext}: during zoom-out, the area around the user’s gaze retains detail, while the periphery zooms out, reducing abrupt changes and helping users maintain context during navigation.}

\textcolor{blue}{\textbf{Medical Imaging:} Radiologists navigate large and detailed images, but traditional input devices such as mice can be inefficient. Alternatives such as head tracking, gesture control, and voice commands have been evaluated for image navigation \cite{headtracking_pathology, gesture_browsing_radiology}. HeadZoom can be adapted for radiological image inspection, using high-resolution chest X-rays obtained, for example, as 12-bit DICOM files from the MIMIC data set; cf. \autoref{fig:application}.b. This use case targets professional radiologists and shifts the task paradigm from target localization to clinical evaluation, similarly to~\cite{sousa17}. Navigation across image sets is achieved using a wireless button, while a Bluetooth mouse allows users to adjust windowing parameters critical for diagnostic visibility, although these tasks could conceivably be performed via speech actions.}

\textcolor{blue}{\textbf{Digital Photography:} Photo browsing is typically based on scrolling grids and pinch-to-zoom gestures. Some researchers investigated alternative approaches such as \textit{PhotoMesa}, a zoomable image browser~\cite{photomesa}. HeadZoom enables users to explore large photo collections using head-based zoom and pan controls. Unlike GIS and medical imaging setups, which prioritize analytical precision, this scenario emphasizes natural navigation and user experience. The interface presents image grids where users can seamlessly zoom in individual photos and pan across the collection cf. \autoref{fig:application}.c, offering a hands-free alternative to traditional gesture-based browsing.}


\section{Design Guidelines}

Based on 
our findings, some guidelines could be provided  for people developing hands-free, controller-free VR interaction, including:

  
  

 
  

\begin{itemize}[nosep,label={},leftmargin=0pt]
    \item \textbf{G1: Minimize effort:} \textcolor{blue}{(From \textit{reduced head movement in Parallel mode}). Parallel mode significantly reduced total head movement, minimizing physical effort/fatigue.}
    \item \textbf{G2: Stabilize input with filtering:} \textcolor{blue}{(Based on our \textit{Kalman filter implementation}). Kalman filters smoothed head tracking, reducing jitter and improving control precision.}
    \item \textbf{G3: Ensure spatial consistency:} \textcolor{blue}{(Justified by \textit{disorientation in Tilt mode}). Tilt mode's dynamic reorientation was disorienting, highlighting the need for predictable image behavior.}
    \item \textbf{G4: Support Visual Anchoring:} \textcolor{blue}{(Following \textit{slightly faster Static mode}). While not significant, Static mode's faster completion times suggest a stable visual anchor aids visual search.} Design interactions that allow users to maintain visual focus during input. Avoid mappings that disrupt line-of-sight or require large shifts in viewpoint.
    \item \textbf{G5: Balance control and simplicity:} \textcolor{blue}{(Justified by \textit{high SUS scores}). All modes achieved high SUS scores ($\geq$ 80), indicating ease of use.}
\end{itemize}

\section{Limitations and Future Research}
HeadZoom advances the field of spatial interaction by enabling fluid, continuous panning and zooming through natural head movement. Although promising, this approach also introduces new challenges and opportunities. HeadZoom demonstrates the feasibility of hands-free head-based interaction for navigating 2D content, but several limitations remain. First, the system relies solely on head pose as input, which constrains the granularity of control and may not scale well to tasks that require high-precision manipulation or sustained focus over extended sessions. In the future it would be good to explore other input alternatives, especially multimodal options, such as combining eye-gaze and head pose.

Another limitation is that the current system assumes stable seating posture and uniform viewing distance, which may not hold in ambulatory or multi-user VR contexts. In the future we should conduct experiments supporting a wider range of physical movement.

Finally, the research was conducted with relatively simple tasks. In order for the results to generalize, future studies should be performed with a broader variety of tasks. This will help establish the applicability of the technique in a wider range of VR settings.



\vspace{0.2cm}
\noindent \textbf{Future Research} should address the following limitations and issues:
    \noindent\textbf{Ergonomics and Fatigue}: Prolonged head movement may cause neck strain or discomfort for the user. Investigating ergonomic thresholds and incorporating adaptive mechanisms could mitigate fatigue and improve long-term usability~\cite{gorillaarm_fatigue}.
    \noindent\textbf{Precision and Responsiveness}: Head-based input must support fine-grained control and minimal latency for effective navigation. Further studies should examine how system responsiveness affects task accuracy and user satisfaction~\cite{gesture_map_interface}.
    \noindent\textbf{Multimodal Integration}: Combining head-based control with complementary modalities, e.g. eye gaze or speech, may improve precision, reduce workload, and support diverse user needs~\cite{eyetracking_survey}.
    \noindent\textbf{Application-Specific Optimization}: Future implementations should tailor head-based navigation to domain-specific requirements, including image granularity in medical imaging, map scale in GIS, and browsing patterns in photography~\cite{headtracking_pathology}.


\section{Conclusions}

This paper presented \textit{HeadZoom}, a hands-free interaction technique that enables users to pan and zoom two-dimensional imagery using natural head motion. Through a controlled user study with three interaction conditions: static, parallel, and tilt, we found that head-based control can be effective in diverse image navigation tasks. 

Among the tested methods, the Parallel mode minimized physical effort while maintaining task performance and was rated most favorably. Statistical analysis confirmed significant reductions in head movement and zoom-related actions, suggesting its suitability for applications requiring sustained or precise visual exploration. Although the Static condition yielded slightly faster average times, it imposed greater physical demand, particularly in rotational metrics. Although promising in theory, the Tilt condition was often perceived as disorienting due to its dynamic image realignment.

Our findings underscore the feasibility of replacing traditional input modalities with controller-free embodied techniques in MR. Future research should explore multi-modal extensions (e.g., gaze, speech) and domain-specific adaptations for clinical, cartographic, and creative workflows. In addition, more research is needed on adaptive calibration strategies and long-term ergonomics to enhance robustness and inclusivity among broader user populations.

\acknowledgments{
The authors wish to thank Anderson Maciel for his comments, the UNESCO Chair on AI \& VR and support from  \textit{Fundação para a Ciência e a Tecnologia} grants
DOI:10.54499/\-UIDB/\-50021/2020,
and 2022.09212.PTDC.}

\clearpage
\bibliographystyle{abbrv-doi}

\bibliography{template}
\end{document}